\documentclass[aps,pra,twocolumn,showpacs,amsmath]{revtex4-1}
\usepackage{graphicx}
\usepackage{amssymb}

\newcommand{\bra}[1]{\langle #1 | \,}
\newcommand{\ket}[1]{\, | #1 \rangle}
\newcommand{\braket}[2]{\langle #1 | #2 \rangle}

\newcommand{\be}{\begin{equation}}
\newcommand{\ee}{\end{equation}}
\newcommand{\bea}{\begin{eqnarray}}
\newcommand{\eea}{\end{eqnarray}}
\newcommand{\besa}{\begin{subeqnarray}}
\newcommand{\eesa}{\end{subeqnarray}}
\newcommand{\bean}{\begin{eqnarray*}}
\newcommand{\eean}{\end{eqnarray*}}

\newcommand{\Nup}{N_{\uparrow}}
\newcommand{\Ndown}{N_{\downarrow}}

\begin{document}

\title{Universal relations for the two-dimensional spin-1/2 Fermi gas with contact interactions}

\author{Manuel Valiente}
\affiliation{Lundbeck Foundation Theoretical Center for Quantum System Research, Department of Physics and Astronomy, Aarhus University, DK-8000 Aarhus C, Denmark}

\author{Nikolaj T. Zinner}
\affiliation{Lundbeck Foundation Theoretical Center for Quantum System Research, Department of Physics and Astronomy, Aarhus University, DK-8000 Aarhus C, Denmark}

\author{Klaus M{\o}lmer}
\affiliation{Lundbeck Foundation Theoretical Center for Quantum System Research, Department of Physics and Astronomy, Aarhus University, DK-8000 Aarhus C, Denmark}

\date{\today}

\begin{abstract}
We present universal relations for a two-dimensional Fermi gas with pairwise contact interactions. The derivation of these relations is made possible by obtaining the explicit form of a generalized function -- selector -- in the momentum representation. The selector implements the short-distance boundary condition between two fermions in a straightforward manner, and leads to simple derivations of the universal relations, in the spirit of Tan's original method for the three-dimensional gas.         

\end{abstract}


\maketitle
%%%%%%%%%%%%%%%%%%%%%%%%%%%%%%%%%%%%%%%%%%%%%%%INTRODUCTION%%%%%%%%%%%%%%%%%%%%%%%%%%%%%%%
\section{Introduction}
The physics of strongly interacting quantum many-body systems has been  
one of the most active fields of research for a number of decades.
Strongly correlated states of matter are ubiquitous in many areas of  
physics ranging from atomic, molecular, and optical \cite{Blochreview},
condensed matter \cite{chaikin} and the study of quantum phase  
transtions \cite{sachdev}, quark-gluon plasma \cite{shuryak}, to the  
physics of neutron stars \cite{pethick}. A prominent example is the
observed non-Fermi liquid behavior in high-temperature superconductor  
materials suggesting a strongly-coupled
state beyond the Landau paradigm \cite{levin}. Perturbative approaches  
to strongly-coupled quantum systems are untenable and even modern  
numerical
techniques can be hard to apply. It is therefore of great interest to  
have analytical insights into systems where interactions are strong.

%The physics of strongly interacting quantum many-body systems has been one of the most active fields of research for a number of decades \cite{FetterWalecka}. These systems, which are relevant in many areas of physics, such as atomic, molecular and optical \cite{BlochReview}, condensed matter \cite{Bruus} and nuclear physics \cite{RingSchuck}, offer challenging theoretical and experimental problems and these attract inmense current interest.

Research in strongly interacting three-dimensional (3D) Fermi gases has experienced an intense growth since the recent appreciation of Tan's universal relations \cite{Tan}. These important results relate many of the many-body properties, such as the adiabatic change of the energy when varying the two-body scattering length, the asymptotic momentum distribution, the two-body loss rate \cite{Braaten} and the pressure to a single quantity called the contact, and were recently verified experimentally \cite{Stewart,Kuhnle}. These universal relations and quantities involved therein have received much recent attention \cite{Nishida,r1,r2,r3,r4,r5,r6,r7,r8,r9,r10}.  

Interestingly, Tan relations have also been obtained for the one-dimensional (1D) Fermi gas \cite{Zwerger}, by means of the operator product expansion \cite{Collins} that had been successfully applied in 3D \cite{Braaten}. However, less is known about these universal relations for the two-dimensional (2D) Fermi gas. Only the energy theorem \cite{Tanarxiv,Combescot,Castin}, which relates the energy of the system to the momentum distribution, the large-momentum tail of the latter \cite{Combescot,Castin} and the adiabatic theorem \cite{Castin} have been so far derived, and only in the homogeneous, or untrapped case. The problem has been recently investigated numerically in ref. \cite{Giorgini}.

In this paper, we derive the universal relations for the zero-range interacting spin-$1/2$ Fermi gas in 2D, with or without an external trapping potential. To do so, we use Tan's original method involving a generalized function, called Tan's selector.  We find the explicit form of the selector in momentum space, in analogy with its 3D counterpart recently obtained by one of us \cite{ValientePseudo}. Our approach greatly simplifies the derivation of the adiabatic theorem with respect to more conventional methods. Once the adiabatic theorem is shown, we use it to derive the generalized virial theorem, the pressure relation and the two-body loss rate.  

\section{System Hamiltonian.}
The zero-range interacting many-body Hamiltonian for $N=N_{\uparrow}+N_{\downarrow}$ spin-$1/2$ fermions has the form
\be
H=\sum_{i=1}^N \frac{\mathbf{p}_i^2}{2m} + \sum_{i<j=1}^N V(\mathbf{r}_{ij})+ \sum_{i=1}^N W(\mathbf{r}_i),\label{Ham}
\ee
where $m$ is the particle's mass, $\mathbf{r}_{ij}=|\mathbf{r}_i-\mathbf{r}_j|$, $W$ is a single-particle external potential, and $V$ is the 2D regularized pseudopotential \cite{Vanzyl}
\be
V(\mathbf{r})=\frac{2\pi\hbar^2}{m}\delta(\mathbf{r}) r\frac{\partial}{\partial r},\label{pseudo}
\ee
which is a particular member of the Olshanii-Pricoupenko $\Lambda$-family of pseudopotentials \cite{Olshanii}. Note that pseudopotential (\ref{pseudo}) does not depend on the 2D scattering length $a$, which only enters the problem through the boundary condition at short-interparticle distances $\psi(\mathbf{r}_{ij})\sim \log(r_{ij}/a)$. The universal singlet binding energy in the homogeneous case ($W=0$) can be seen to be \cite{verhaar}
\be
E_B = \frac{4\hbar^2 e^{-2\gamma}}{m a^2},\label{binding}
\ee
where $\gamma = 0.577215665...$ is Euler's constant. The binding energy is the most appropriate fitting parameter of the theory in momentum space, as we see in the following paragraph.

\section{Tan's selector.}
To derive the universal relations for the Fermi gas in 2D, we will make use of Tan's original method \cite{Tan}, particularized to two dimensions. In ref. \cite{Tanarxiv}, Tan defined two selectors, $\tilde{\eta}$ and $l$, the last of which is unnecessary in 2D, and involves the logarithm of a dimensioned quantity. The $\tilde{\eta}$-selector satisfies $\int \mathrm{d}^2r\tilde{\eta}(r)=1$, $\int \mathrm{d}^2r\tilde{\eta}(r)\log(r/a)=0$, while $\tilde{\eta}(r)=0$ for $r\ne 0$. 

We here redefine the momentum representation $\eta(k)$ of $\tilde{\eta}(r)$ as follows
\begin{align}
&\eta(k)=1 \hspace{0.1cm} (k<\infty),\\ 
&\int\mathrm{d}^2k \eta(k) \frac{1}{\hbar^2 k^2/m + E_B}=0, \label{condition2}
\end{align}
where $E_B$ is the two-body universal binding energy, Eq. (\ref{binding}).
It is easy to see that Eq. (\ref{condition2}) is equivalent to Eq. (16b) of ref. \cite{Tanarxiv}, but our condition is more natural and represents a clear way of writing the two-body bound state integral equation. Proceeding in analogy to the 3D case \cite{ValientePseudo}, we see that 
\be
\eta(k)=1+\log\left(\frac{2e^{-\gamma}}{ka}\right)\frac{\delta(1/k)}{k}.\label{etaexplicit}
\ee
The above explicit form for $\eta(k)$ will prove crucial for the derivation of the two-dimensional adiabatic theorem. 

\section{Energy theorem.}
In ref. \cite{Tanarxiv} Tan stated, without derivation, that the energy of the 2D Fermi gas can be written as
\be
E=\sum_{\mathbf{k}\sigma} \eta(k) \frac{\hbar^2 k^2}{2m} n_{\mathbf{k}\sigma} + \langle \mathcal{W} \rangle,\label{energyrelation}
\ee
where $n_{\mathbf{k}\sigma}=\langle c_{\mathbf{k},\sigma}^{\dagger}c_{\mathbf{k},\sigma}\rangle$, with $c_{\mathbf{k},\sigma}$ the spin-$\sigma$ annihilation operator at momentum $\mathbf{k}$, is the momentum distribution and $\mathcal{W}=\sum W(\mathbf{r}_i)$ is the total external potential.  

To prove Eq. (\ref{energyrelation}) it is sufficient to consider a pure, normalized state $\ket{\phi}$ with $N_{\sigma}$ fermions of spin $\sigma=\uparrow, \downarrow$ and total number of fermions $N=\Nup+\Ndown$
\be
\ket{\phi}= \frac{1}{\Nup ! \Ndown !} \int \mathrm{d}\mathbf{R}\mathrm{d}\mathbf{S} \phi(\mathbf{R},\mathbf{S}) \prod_{i=1}^{\Nup} \psi_{\uparrow}^{\dagger}(\mathbf{r}_i)   \prod_{j=1}^{\Ndown} \psi_{\downarrow}^{\dagger}(\mathbf{s}_j)\ket{0},
\ee
where $\mathbf{R}$ ($\mathbf{S}$) is shorthand for $\mathbf{r}_1,\ldots,\mathbf{r}_{\Nup}$ ($\mathbf{s}_1,\ldots,\mathbf{s}_{\Ndown}$) and the integrals are done over the whole $2N$-dimensional space. Above, $\psi_{\sigma}^{\dagger}(\mathbf{r})$ is the spin-$\sigma$ creation operator at position $\mathbf{r}$, and $\ket{0}$ is the vacuum.

Tan's derivation \cite{Tan} for the 3D gas can be followed in parallel for the 2D case until the following expression is encountered
\be
\int \mathrm{d}\mathbf{R}'\mathrm{d}^2r_0 \int \mathrm{d}^2t \tilde{\eta}(t) \nabla_{\mathbf{t}}^2 K(\mathbf{R}',\mathbf{r}_0,\mathbf{t}), \label{integral1}
\ee
where $\mathbf{R}'=(\mathbf{r}_2,\ldots,\mathbf{r}_{\Nup},\mathbf{s}_2,\ldots,\mathbf{s}_{\Ndown})$, $\mathbf{r}=\mathbf{r}_1-\mathbf{s}_1$ and $\mathbf{r}_0=(\mathbf{r}_1+\mathbf{s}_1)/2$, and $K$ is defined as
\be
K(\mathbf{R}',\mathbf{r}_0,\mathbf{t})=\int_{r<\epsilon} \mathrm{d}^2r\phi^*(\mathbf{R}',\mathbf{r}_0,\mathbf{r})\phi(\mathbf{R}',\mathbf{r}_0+\mathbf{t}/2,\mathbf{r}+\mathbf{t}),\label{Kdef}
\ee
where $\phi(\mathbf{R}',\mathbf{r}_0,\mathbf{r})$ stands for $\phi(\mathbf{R},\mathbf{S})$, with $\mathbf{r}_1$ and $\mathbf{s}_1$ replaced by $\mathbf{r}_0+\mathbf{r}/2$ and $\mathbf{r}_0-\mathbf{r}/2$, respectively. In Eq. (\ref{Kdef}), $\epsilon>0$ is a small real number, however much larger than $t=|\mathbf{t}|$. The goal is to show that the integral in Eq. (\ref{integral1}) vanishes since, if it does, relation (\ref{energyrelation}) holds \cite{Tan}. First, we expand $\phi$  as 
\be
\phi(\mathbf{R}',\mathbf{r}_0,\mathbf{r})= A(\mathbf{R}',\mathbf{r}_0)[\log(r/a)+O(r)],\label{phiexp}
\ee
and 
\be
A(\mathbf{R}',\mathbf{r}_0+\mathbf{t}/2) = A(\mathbf{R}',\mathbf{r}_0)+\nabla_{\mathbf{r}_0}A(\mathbf{R}',\mathbf{r}_0)\cdot \mathbf{t}/2+O(t^2).
\ee
We obtain $K=K_0+K_1$, with 
\begin{align}
&K_0\approx \pi|A(\mathbf{R}',\mathbf{r}_0)|^2 \frac{t^2}{2} [\log(t/a)-1] \\ 
&K_1\approx\pi A^*(\mathbf{R}',\mathbf{r}_0)\frac{t^3}{4}(\log(t/a)-1)\nabla_{\mathbf{r}_0}A(\mathbf{R}',\mathbf{r}_0)\cdot \hat{\mathbf{t}}.
\end{align}
Inserting $K$ into (\ref{integral1}), we see that the integral vanishes, as we wanted to show.

We can express the energy relation in Eq. (\ref{energyrelation}) more explicitly by using the form of the $\eta$-selector in Eq. (\ref{etaexplicit}), as
\be
E=\frac{\hbar^2\log(2e^{-\gamma})}{2\pi m} \Omega C + \sum_{\mathbf{k},\sigma} \frac{\hbar^2 k^2}{2m}\left[n_{\mathbf{k},\sigma} - \frac{C}{k^3(k+a^{-1})}\right]+ \langle \mathcal{W} \rangle,\label{energyrelationexplicit}
\ee
where $\Omega$ is the area and $C=\lim_{\mathbf{k}\to \infty} k^4 n_{\mathbf{k},\sigma}$ is the 2D contact density. The above relation, Eq. (\ref{energyrelationexplicit}), was expressed in a somewhat different manner by Combescot {\it et al.} in \cite{Combescot} and by Werner and Castin in \cite{Castin}.  
The tail of the momentum distribution was found in \cite{Combescot,Castin} to be $\propto 1/k^4$ and therefore the contact $C$ is a finite quantity.

\section{Adiabatic theorem.}
The adiabatic theorem states the relation between the contact and the change in energy as the scattering length is slightly varied. This is the central result among the universal relations for the Fermi gas, since it is needed to derive the virial theorem, the pressure relation and the inelastic two-body loss rate. 

The fact that the 2D pseudopotential, Eq. (\ref{pseudo}), does not depend explicitly on the scattering length adds a complication to the proof of the adiabatic theorem that was not present in 3D and 1D. Indeed, the Hellmann-Feynmann theorem, which was employed to derive this important result in 3D by Braaten and Platter \cite{Braaten} and in 1D by Punk and Zwerger \cite{Zwerger}, cannot be applied directly to Eq. (\ref{pseudo}) nor to any member of the so-called $\Lambda$-family of pseudopotentials \cite{Olshanii,Vanzyl}. In addition, Tan's approach to the 2D problem \cite{Tanarxiv} lacked a simple dependence of the $\eta$-selector on the scattering length. These two facts have prevented a simple derivation of the adiabatic theorem using regularized pseudopotentials, which has been first shown by Werner and Castin \cite{Castin} who obtained it using Bethe-Peierls boundary condition. Below, we present a very simple proof of the adiabatic theorem by using the explicit form, Eq. (\ref{etaexplicit}), of the $\eta$-selector. 

We begin by defining the following operator
\be
\eta_aT\equiv \sum_{\mathbf{k},\sigma}\eta(k) (\hbar^2 k^2/2m)c_{\mathbf{k},\sigma}^{\dagger}c_{\mathbf{k},\sigma},
\ee
where $\eta(k)$ corresponds to scattering length $a$. It is then easy to see that
\be
E(a')-E(a) =\frac{\bra{\phi(a)}\eta_aT-\eta_{a'}T\ket{\phi(a')}}{\braket{\phi(a)}{\phi(a')}},
\ee
where $E(a)$ and $\phi(a)$ are an energy eigenvalue and associated eigenstate of Hamiltonian (\ref{Ham}) corresponding to scattering length $a$. Using Eq. (\ref{etaexplicit}) we find 
\be
\frac{E(a')-E(a)}{\log(a'/a)} =\sum_{\mathbf{k},\sigma} \frac{\hbar^2k^2 \delta(1/k)}{2mk} \frac{\bra{\phi(a)} c_{\mathbf{k},\sigma}^{\dagger}c_{\mathbf{k},\sigma} \ket{\phi(a')}}{\braket{\phi(a)}{\phi(a')}}.
\ee
If we now take the limit $a'\to a$ in the above equation, we obtain the desired adiabatic theorem
\be
\frac{dE}{da}(a) = \frac{\hbar^2\Omega}{2\pi m a} C. \label{adiabatic}
\ee

An immediate application of the adiabatic theorem is the calculation of the contact for a weakly-coupled 2D Fermi gas. The energy density for the spin-balanced case is given by \cite{Energy2D}
\be
\frac{E}{N}\approx \frac{\hbar^2 k_F^2}{4m}\left(1-\frac{1}{\log(k_Fa)}\right),
\ee
where $N$ is the number of particles and $k_F$ is the Fermi momentum. The contact in this limit is therefore given by
\be
C=\frac{\pi \rho k_F^2}{2\log^2(k_F a)}.
\ee
where $\rho=N/\Omega$ is the particle density.

\section{Generalized virial theorem.}
We assume now that the system is under an external potential of the form $W(\mathbf{r}) \propto r^{\beta}$, which has not been considered before for the 2D case. In this case, we will show that the virial theorem reads
\be
E= \frac{\beta +2}{2} \langle \mathcal{W} \rangle -\frac{\hbar^2 \Omega C}{4\pi  m}.\label{virial}
\ee
To see this, we follow a technique used for the 3D gas \cite{Tan}. We begin with a state $\phi$ corresponding to scattering length $a$. We slightly change the scattering length to $a'=(1+\epsilon)a$, with $\epsilon$ small, associated with state $\phi'$ which has energy $E'$. Expanding $E'$ in powers of $\epsilon$, and using the adiabatic theorem, Eq. (\ref{adiabatic}), we find 
\be
E'=E+\epsilon \frac{\hbar^2\Omega C}{2\pi m} + O(\epsilon^2).\label{equationtwice}
\ee
We now define a rescaled wave function $\phi''(\mathbf{R},\mathbf{S}) \equiv (1+\epsilon)^N \phi'((1+\epsilon)\mathbf{R},(1+\epsilon)\mathbf{S})$. Expanding $\phi''$ as in Eq. (\ref{phiexp}), we find that $\phi''$ is a state at scattering length $a''=a'/(1+\epsilon)=a$. Using the scaled wave function in the expectation value of the energy, we find its energy $E''$ to be
\be
E''=(1+\epsilon)^2E_{\mathrm{in}}'+(1+\epsilon)^{-\beta}\bra{\phi'}\mathcal{W}\ket{\phi'},
\ee
where $E_{\mathrm{in}}'=E'-\bra{\phi'}\mathcal{W}\ket{\phi'}$. Expanding $E''$ in powers of $\epsilon$, we obtain $E''-E'=2\epsilon E_{\mathrm{in}}' - \beta \epsilon \bra{\phi'}\mathcal{W}\ket{\phi'} + O(\epsilon^2)$. From the quadratic convergence properties of variational energies \cite{Porras} we have $E''-E= O(\epsilon^2)$, and therefore, using Eq. (\ref{equationtwice}) twice, we obtain 
\be
2\epsilon E_{\mathrm{in}}-\beta \epsilon\bra{\phi}\mathcal{W}\ket{\phi} + \epsilon \frac{\hbar^2 \Omega C}{2\pi m} = O(\epsilon^2),
\ee
which, after taking the limit $\epsilon \to 0$, proves the virial theorem (\ref{virial}).  

\section{Pressure relation.}
For a homogeneous system ($W=0$), there is a relation between pressure, energy and contact, which has been derived in 3D \cite{Tan,Braaten} and 1D \cite{Zwerger}. In 2D, it is given by 
\be
P= \frac{E}{\Omega}+ \frac{\hbar^2 C}{4 \pi m},\label{pressure}
\ee
To show Eq. (\ref{pressure}), we follow most of the steps taken for the proof of the virial theorem (\ref{virial}), and we arrive at the expression
\begin{align}
P\Omega&=\lim_{\epsilon \to 0} \frac{E''-E}{1-(1+\epsilon)^{-2}}\nonumber \\
&=\left(2E+\frac{\hbar^2\Omega C}{2\pi m}\right)\lim_{\epsilon \to 0} \frac{\epsilon}{1-(1+\epsilon)^{-2}}
\end{align}
Using L'h{\^o}pital's rule in the above equation, we obtain the desired pressure relation (\ref{pressure}). 

\section{Inelastic two-body loss rate.}
Two-particle loss rates can be calculated \cite{Braaten} by adding a small imaginary part $a_I$ ($|a_I|\ll |a|$) to the scattering length, so that we replace $a\to a+\mathrm{i}a_I$. The resulting complex energy is then expanded as $E(a+\mathrm{i}a_I) = E(a)+\mathrm{i}a_I dE/da(a)+O((a_I)^2)$, and the inelastic loss rate $\Gamma$ is identified as $-\Gamma/2=a_IdE/da$. Using the adiabatic theorem (\ref{adiabatic}), we obtain
\be
\Gamma = -a_I\frac{\hbar^2 \Omega C}{\pi m a} + O((a_I)^2).
\ee  

\section{Conclusions.}
We have obtained the universal Tan relations for the two-dimensional spin-$1/2$ Fermi gas with zero-range interparticle interactions. These results can be tested with current experimental techniques such as Bragg \cite{Kuhnle} or RF spectroscopy \cite{Stewart,Zwerger2,Langmack}, already available for the two-dimensional Fermi gas \cite{Koehl}. Our methodology is close to the original approach for the three-dimensional problem \cite{Tan}. The power of our approach relies on obtaining the explicit form, in momentum representation, of the so-called Tan's selector, which expresses the contact conditions between two particles with a given scattering length in closed form. This allowed us to overcome the fact that the Hamiltonian of the system does not depend on the two-body scattering length, and led us to a more straightforward derivation than with conventional methods.  
Our method is directly applicable also to universal three-body physics  
in two dimensions \cite{tjon,Hammer,bellotti} where
additional three-body contact parameters were recently proposed in the  
three dimensional case \cite{braaten2011,castin2011}. This is an  
interesting direction
for future study.

%%%%%%%%%%%%%%%%%%%%%THANKS%%%%%%%%%%%%%%%%%%%%%%
\begin{acknowledgments} 
MV acknowledges support from a Villum Kann Rassmussen block Scholarship. NTZ is supported by the Danish Council for Independent Research $|$ Natural Sciences.
\end{acknowledgments}

\end{document}